\documentstyle[preprint,eqsecnum,aps]{revtex}
\tightenlines

\newcommand{\cs}{\'{c}}
\newcommand{\ch}{\v{c}}

\newcommand{\beq}{\begin{equation}}
\newcommand{\eeq}{\end{equation}}
\newcommand{\bdm}{\begin{displaymath}}
\newcommand{\edm}{\end{displaymath}}
\newcommand{\beqa}{\begin{eqnarray}}
\newcommand{\eeqa}{\end{eqnarray}}
\newcommand{\beqab}{\begin{eqnarray*}}
\newcommand{\eeqab}{\end{eqnarray*}}

 \def\@makefnmark{\hbox to 0pt{$^{\@thefnmark}$\hss}}  

\newcounter{saveeqn}%

\begin{document}

\draft
\preprint{}

\title{On Bethe strings in the two-particle sector of the closed $SU(2)_{q}$ invariant spin chain}  
\author{Amon Ilakovac \footnote{e-mail: ailakov@phy.hr},
Marko Kolanovi\cs \footnote{e-mail: kmarko@phy.hr},
Silvio Pallua  \footnote{e-mail: pallua@phy.hr} and
Predrag Prester \footnote{e-mail: pprester@phy.hr} }
\address{Department of Theoretical Physics, University of Zagreb\\
Bijeni\ch ka c.32, POB 162, 10001 Zagreb, Croatia}
\date{\today}
\maketitle
\begin{abstract}
In this paper we investigate complex solutions of the Bethe equations in the two-particle
sector both for arbitrary finite number of sites and for the thermodynamic limit
. We find the number of complex solutions (strings) and 
compare it with the string conjecture prediction. Some simple properties of these solutions
like position in the spectrum, crossing of levels, connection to the ground state and 
transformation to the real solutions are discussed. Counting both real and complex solutions
we find expected number of highest weight Bethe states.
\end{abstract}

\narrowtext

\section{Introduction}

Among integrable spin chains those invariant on quantum groups   have attracted 
 considerable interest . The simplest among these are  $SU(2)_{q}$ invariant chains. 
Open chains have been considered for spin one half \cite{A1}, spin one \cite{D5} and higher spin
 \cite{D0}.
Generalisations to other groups have been also investigated \cite{batc}, \cite{ritt}.
 Closed chains have been introduced later because parallel requirement
of quantum group invariance and generalised translational invariance required introduction
of a nonlocal term in the Hamiltonian \cite{A2}. These chains were shown to have interesting properties. 
$SU(2)_{q}$ invariant closed chain has ground state with vanishing or nonvanishing spin 
depending on the value of the coupling constant \cite{A0}. Central charge was found and it 
was shown that in particular points of the coupling constant
 it corresponds to central charge of minimal unitary series \cite{A0}, \cite{karo}.
Its excited states and operator content was also found \cite{pp}. Recently, it was 
argued that this model is related to interesting field theories \cite{fior} and in 
particular to the Liouville theory with imaginary coupling \cite{fade}. A common approach
to investigation of integrable models in general and of this model in particular is the Bethe
ansatz method \cite{C4}, \cite{C5}. This method leads to a set of transcendental equations for momenta
of quasiparticles. One class of the solutions has real quasimomenta that can be found by numerical 
iteration. However, there are also solutions with complex quasimomenta. Searching for the latter class
of solutions one usually makes the so-called string conjecture \cite{C6}, \cite{C7}. In particular, it has
been used in investigations of some quantum invariant models \cite{C15}, \cite{C16} and also in  a recent
investigation motivated by the relation of such models to Liouville field theories  \cite{nasa}.
However, it was recently pointed  in the context of XXX and XXZ models that string conjecture has 
exceptions \cite{C8,C9,C10,C11,C12,C13,C0}. 
That means those solutions to Bethe Ansatz equations
are not yet completely understood. For that reason in this paper 
we investigate  complex solutions for the closed quantum invariant spin chain. In this 
investigation we shall not use the string conjecture.        

For simplicity, we shall investigate the sector  where spin is lower by two units than 
the maximal spin. We shall find the number of complex solutions for arbitrary number
of sites $N$. In the thermodynamic limit, this number is found to be of the order $N$, as expected on the 
basis of the string conjecture. Next correction will differ for a finite number from the
string conjecture prediction. If the coupling constant tends to the Bethe point $(\cos\phi\rightarrow 1)$
in a sufficiently fast manner, than we shall have an infinite number of exceptions to the number 
predicted by the string conjecture. The model in this point coincides with the XXX model and 
this is consistent with results of \cite{C11}, \cite{C12}. Further, we shall find  that the energy distribution 
of complex solutions show simple features and that they are located in 
a narrow energy band. In particular they are located on the top of the spectrum near the 
antiferromagnetic point and on the bottom near the ferromagnetic point. 
Energy levels cross each other as coupling constant changes. There are no such crossings in 
the XXZ model. We also find that  solutions of the one of the two classes of bound states (strings)
evolve in real solutions in the special points for the coupling constant where we know 
that the representation theory is not isomorphic to usual $SU(2)$. It will turn also that at
least one of overall ground states evolves in the string state. This happens in a region of
coupling constant where it is not more ground state.
      
As already anticipated we shall consider the Hamiltonian
\beq \label{1}
H=Nq - \sum_{i=1}^{N-1} R_{i}-R_{0}
\eeq
 \beq \label{2}
R_{0} = G R_{N-1} G^{-1}
\eeq
\beq \label{3}
G=R_{1} R_{2}....R_{N-1}
\eeq
where $R_{i} $ are $4\times 4$ matrices given with
\beq  \label{4}
R_{i} =\sigma_{i}^{+} \sigma_{i+1}^{-} + 
\sigma_{i}^{-} \sigma_{i+1}^{+} +  
\frac{q+q^{-1}}{2} 
 \left ( \sigma_{i} ^{3} \sigma_{i+1}^{3} + 1 \right ) - 
 \frac{q-q^{-1}}{4} \left (  \sigma_{i} ^{3}
- \sigma_{i+1}^{3} - 2 \right ) .    
\eeq
Here $q$ denotes a parameter which lies on unit circle
\beq \label{5}
q=e^{i\phi}
\eeq
and $\phi$ will be called coupling constant.
The operators $R_{i}$ satisfy Hecke algebra
\beq \label{6}
R_{i}^{2} = \left ( q-q^{-1} \right ) R_{i}  +  1
\eeq
\beq \label{7}   
R_{i} R_{i+1} R_{i}  =  R_{i+1} R_{i} R_{i+1}.
\eeq
As a consequence
\beq \label{8}
             [G,H]=0.
\eeq
In addition it was shown that this Hamiltonian is invariant on the
$SU(2)_{q}$ symmetry group whose generators are given with
\beq \label{9}
S^{3}  =\frac{1}{2}  \sum_{i=1}^{L} 1\otimes \ldots\otimes\sigma_{i}^{3}
 \ldots\otimes 1
 \eeq               
\beq \label{10}
  S^{+}  =  \sum_{i=1}^{L} q^{-\sigma^{3}/2} \otimes.. ..
\sigma_{i}^{+} \otimes.......\otimes q^{\sigma^{3}/2}. 
\eeq
\beq \label{11}
    [S^{+} ,S^{-}] =  \frac{q^{2S^{3}}-q^{-2S^{3}}}{q-q^{-1}}.
\eeq                    
Hamiltonian of the model is highly nonlocal but due to Hecke algebra it is still integrable. 
It can be diagonalized e.g. with the coordinate Bethe Ansatz method \cite{A0}.
The energy eigenfunctions of the spin $s$
\beq \label{12}
    s=\frac {N}{2}-M,
\eeq
with $M$ spins down can be written as
\beq \label{13}
\ |\psi_{M}\rangle =  \sum_{1\leq n_{1} \leq n_{2} \ldots \leq n_{M} \leq N}
\psi_{M}\left (n_{1},\ldots n_{M}\right ) |n_{1}\ldots n_{M}\rangle.
\eeq

The $\psi_{M}$ functions are given with
\beq \label{14}
\psi_{M}( n_{1},\ldots n_{M})  =  
\sum_{P} \exp  \left [ i \left ( \sum_{j=1}^{M}k_{P_{j}}n_{j} + 
\frac{1}{2} \sum_{1 \leq j \leq l \leq M} \Phi_{P_{j},P_{l}}  \right )  \right ].
\eeq   
Here the sum runs over the elements of the permutation group $S_{M}$.
The phase shifts $\Phi_{j,i}$ have the following simple expression

\beq \label{phi}
 \Phi_{j,i}=2\arctan \frac{\cos\phi\sin\frac{(k_{j}-k_{i})}{2}}{\cos\frac{(k_{j}+k_{i})}{2}
 - \cos\phi\cos \frac{(k_{j}-k_{i})}{2}}.
\eeq
The quasimomenta $k_{i}$, $i=1 \ldots M$ form a solution of the Bethe 
Ansatz equations 
\beq \label{bet}
Nk_{i} + \phi \left ( 2M-N-2 \right ) + \sum_{j=1}^{M}
\Phi_{i,j}= 2\pi \lambda_{i} \quad i=1\ldots M.
\eeq
The $M$ Bethe numbers $I_{i},i=1 \ldots M$ are half integers (integers)
for $M$ even (odd). In terms of quasimomenta $k_{i}, i=1 \ldots M$ the energy
$E$ and generalized momentum $P$ read
\beq \label{17}
E = 2 \sum_{i=1}^{M} ( \cos \phi  - \cos k_{i} ),
\eeq
\beq \label{18}
P = \sum_{i=1}^{M} k_{i}  - \phi (N -M-1).
\eeq
Operator $G$ is then 
\beq \label{19}
G = e^{-iP}.
\eeq
Due to $SU(2)_{q}$ symmetry one can use its representation theory
to classify the states. In fact for generic $q$ we have the same 
multiplet  structure as for usual $SU(2)$. However, for
\beq \label{20}
 q^{p} = \pm 1~~~~~~~~~~~~~ p~~~~ \mbox{integer} 
\eeq
additional degeneracy occur \cite{A1}, \cite{A2}. In particular representations
of spins  $j ^{'}=j+np$ and $ j^{'}=p-1-j-np$ mix. Here $n$ is an integer. In order to 
get representations isomorphic to $SU(2)$  one has to exclude 
also
\beq \label{22}
j=np - 1/2.
\eeq
Only the remaining representations (called 'good' representations) with
\beq \label{23}
j<\frac{p-1}{2}         
\eeq
are isomorphic to $SU(2)$. The parameter $q$ which is a root of unity can be written as
\beq \label{24}
q = e^{ i \frac{ \pi p'}{p}}.
\eeq
It was shown \cite{A1} that for $ p'=1$ the representations are unitary.
We shall find afterwards that the points defined with (\ref{24}) will play a 
role in the evolution of complex
solutions into real ones when we vary the coupling constant. Let us for the moment concentrate
to the generic sector of the parameter $q$. Due to the already mentioned property that
in this case we have the usual $SU(2)$ multiplet structure it is sufficient to find the highest
weight states. Other states can be constructed \cite{A0} by application of the generator $J^{-}$.
It is known that the highest weight states correspond to sets of $\{k^{i}\},i=1\ldots M$ where 
the quasimomenta
satisfy
\beq \label{25}
k_{i} \neq  \phi.
\eeq
In fact we can identify in advance Bethe numbers which lead to the non-highest weight solutions.
Let us assume that for some $i=l$ we have
\beq \label{26}
k_{l} = \phi.
\eeq
A straightforward calculation using (\ref{phi}) shows that
\beq \label{27}
\Phi_{l,j}\left (k_{l}=\phi,k_{j}\right )=\pi-2\phi.
\eeq 
From Bethe equation (\ref{bet}) for $ i=l$ we obtain
\beq \label{28}
 \lambda_{l}=\frac{M-1}{2}
\eeq
The highest weight solutions will be obtained by excluding
(\ref{28}) from the choice of the Bethe numbers. The ground state for the Hamiltonian (\ref{1})
was found for the whole interval $ 0 \leq \phi \leq \pi$. In fact, contrary to the the XXZ 
model, the spin zero state is the ground state only in the $ \frac{\pi}{2} \leq \phi \leq \pi $ region.
In the rest of the interval there is a subregion for each spin  where the ground state has 
just that particular spin. More precisely, the total spin $s$ of the ground state depends on the coupling
constant $\phi $ according to
\beqa \label{ps}
  J=0\quad \mbox{for}\quad \frac{\pi}{2}\leq \phi \leq \pi  \nonumber \\
 \label{29}  J=s\quad\mbox{for}\quad\frac{\pi}{2(s+1)} \leq\phi\leq 
\frac{\pi}{2s}  \\
  J=\frac{N}{2}~~\mbox{for}~~0 \leq \phi \leq \frac {\pi}{N}.
 \nonumber
  \eeqa
The Bethe numbers which give ground state are given with
\beq \label{30}
 \lambda=\left ( -\frac{M-1}{2} \right ) - 1, \ldots , \left (\frac{M-1}{2} \right )
 - 1.
\eeq
and
\beq \label{31}
M=\frac{N}{2} - s.
\eeq
It also turns out that the corresponding momenta are real. The natural question arises which
role play complex solutions of Bethe equations.

\section{Complex solutions in two particle sector}
\setcounter{equation}{0}

In this section we shall investigate complex solutions of the Bethe equations (\ref{bet}) without
assuming the string conjecture \cite{C6},\cite{C7}. For simplicity we shall work in the $M=2$ sector. In 
that case the Bethe equations (\ref{bet}) take the form

\beqa\label{tps}
Nk_{1}+\Phi_{1,2}+\phi(2-N)=2\pi\lambda_{1},\\
\label{tps1} Nk_{2}-\Phi_{1,2}+\phi(2-N)=2\pi\lambda_{2}.
\eeqa
Here we want in particular to look for complex solutions. Due to reality of energy
and generalized momentum, $k_{1}$ and $k_{2}$ have to be complex conjugates of each other
\beqa \label{k1k2}
k_{1}=k_{r}+ik_{i},\\
k_{2}=k_{r}-ik_{i}.
\eeqa
We can express $k_{r}$ and $k_{i}$ by taking sum and difference of equations
(\ref{tps}) and (\ref{tps1})
\beq \label{kr}
k_{r}=\frac{1}{N}\left[\pi(\lambda_{1}+\lambda_{2})-\phi(2-N)\right]
\eeq
\beq \label{ki}
iNk_{i}=\pi(\lambda_{1}-\lambda_{2})-2\arctan\frac{\cos\phi\sin(ik_{i})}
{\cos k_{r}-\cos\phi\cos(ik_{i})}.
\eeq
With the help of the identity
\beq \label{arctan}
\arctan z=\frac{1}{2i}\ln\frac{1+iz}{1-iz}
\eeq
and exponentiation of (\ref{ki}) one can obtain

\beq \label{sh}
\frac{\sinh (k_{i}(\frac{N}{2}-1))}{\sinh (k_{i}\frac{N}{2})}=\frac{\cos k_{r}}{\cos\phi}
\qquad\lambda_{1}+\lambda_{2}\quad \mbox{odd},
\eeq

\beq \label{ch}
\frac{\cosh (k_{i}(\frac{N}{2}-1))}{\cosh (k_{i}\frac{N}{2})}=\frac{\cos k_{r}}{\cos\phi}
\qquad\lambda_{1}+\lambda_{2}\quad \mbox{even}.
\eeq
In the further analysis we shall call the solutions of equation (\ref{sh}) s-solutions and the
solutions of equation (\ref{ch}) c-solutions. 
The left hand sides of both equations (\ref{sh}), (\ref{ch}) are monotonously decreasing functions 
so we shall have a solution for $k_{i}$ for any $k_{r}$ for which $\cos k_{r}$ is in the interval
\beqa \label{int1}
\mbox{s-strings}&:&\quad 0\leq\cos k_{r}<\cos\phi(1-\frac{2}{N}),
\\ \mbox{c-strings}&:&\quad 0\leq\cos k_{r}<\cos\phi
\eeqa  
if $\cos\phi\ge 0\quad (0\leq\phi\le\frac{\pi}{2})$ and
\beqa \label{int2}
\mbox{s-strings}&:&\quad \cos\phi(1-\frac{2}{N})<\cos k_{r}\leq 0,
\\ \mbox{c-strings}&:&\quad \cos\phi<\cos k_{r}\leq 0
\eeqa
if $\cos\phi\le 0 \quad(\frac{\pi}{2}\leq\phi\le\pi)$. Further, as long as $k_{i}\neq 0$,
$k_{1,2}\neq\phi$ and thus this solution represents the highest weight state. Now we can 
proceed to find the number of complex solutions. As a first step we shall  consider leading order 
in $N$ when the inequalities (\ref{int1}) and (\ref{int2}) are identical. In this case, for a given 
coupling constant $\phi$, the number of complex solutions will depend on the number of values that $k_{r}$
can take.  In order to have one to one 
correspondence between $k_{r}$ and $\cos k_{r}$, the sum of Bethe numbers can take $2N-1$ different
 equidistant values . The interval in $k_{r}$ for which there are 
complex solutions is $2\left(\frac{\pi}{2}-\phi\right)$. As a result the number of the string
solutions in leading order in $N$ is
\beq \label{nta}
\frac{1}{2\pi}(2N-1)(\pi-2\phi).
\eeq
Maybe we shall remark that if we insist to have $k_{r}$ between $\pi$ and $-\pi$ we should not
take Bethe numbers symmetrically around zero due to the term proportional to $\phi$ in equation
(\ref{kr}). However, this does not affect the counting argument. To determine the number
of complex solutions more precisely, we have to take into account subleading orders in $N$ in relations 
(\ref{int1}) and (\ref{int2}). The allowed interval for the real parts of s-solutions is smaller than
the corresponding interval for c-solutions and there are no s-solutions for an interval in $k_{r}$ of length
\beq \label{dk}
\delta k_{r}=4\arcsin
\frac{\cos\phi}{N\sin\left(\frac{\arccos\cos\phi+\arccos\left(\cos\phi(1-\frac{2}{N})\right)}{2}\right)}.
\eeq
The number of solutions, which were overcounted in naive formula (\ref{nta}), is integer part of
\beq \label{nnn}
n=\frac{2N-1}{\pi}\arcsin\frac{\cos\phi}{N\sin\left(\frac{\arccos\cos\phi+\arccos\left(\cos\phi(1-\frac{2}
{N})\right)}{2}\right)}.
\eeq
This correction is a finite number even in the thermodynamic limit $(N\rightarrow\infty)$.
However, if $|\cos\phi|=1$ correction is infinite and goes as $\sqrt{N}$. This is true if
$|\cos\phi|\rightarrow 1$ sufficiently fast, more precisely if
\beq \label{uvj}
N(1-|\cos\phi|)^{\alpha}=\mbox{const.}\quad\mbox{and}\quad\alpha>1.
\eeq
This is consistent with the fact that $\phi=0$ or $\phi=pi$ corresponds to XXX chain for which is
known that this correction is infinite. This correction represents at the same time 
a violation of the string conjecture which was found previously in XXX \cite{C11}, \cite{C12} 
and XXZ chain \cite{C13}, \cite{C0}. 
Disappearance of s-solutions (strings) is followed by appearance of real solution
with two close quasimomenta. These quasimomenta can be found near the number of sites
where s-string disappeared by numerical iteration of (\ref{tps})
and (\ref{tps1}) with the same Bethe numbers $\lambda_{1}$ and $\lambda_{2}$. Real solutions of
Bethe equations with two identical Bethe numbers also represent violation of the string
conjecture. Evolution of s-strings into real solutions with identical Bethe numbers 
can be followed for fixed coupling constant
by increasing the number of sites $N$. Exapmle is given on Fig.1 for $\phi=0.32$.

\section{Properties of complex solutions}
\setcounter{equation}{0}

Next interesting question we would like to ask is how are complex solutions (bound states)
distributed on the energy scale. From the relations (\ref{ch}), (\ref{sh}) and (\ref{17}) we 
see that complex solutions are confined in the energy band
\beqa \label{e2}
0< E(\mbox{c-strings})\leq 2\cos\phi,\\
\frac{8\cos\phi}{N}< E(\mbox{s-strings})\leq 2\cos\phi.
\eeqa
The left sides of the inequalities correspond to the points (in coupling constant $\phi$) where two 
complex quasimomenta collide on the real axis and the complex solution becomes real solution
(decay of the bound state). The right sides  of the inequalities correspond to the points where 
$k_{i}$ tends to infinity and so the localisation of two overturned spins in bound state
tends to infinity. This can be seen from the form of the Bethe wave function (\ref{14}). 
The energy band (\ref{e2}) is generally narrow compared to the overall spread of energy when
all solutions are included. As an illustration of $M=2$ spectrum, figures are given for 
$N=6, N=8$ and $N=10$ (Fig.2, Fig.3, Fig.4). We see that all bound states (strings) 
disappear near the 'free theory' point $\phi=\frac{\pi}{2}$.
Energies of the string solutions are on the top
of the spectrum near the Bethe antiferromagnetic point $\phi=\pi$ and on the bottom of the 
spectrum near the Bethe ferromagnetic point $\phi=0$. There is an interesting question 
connected to the nature of the overall ground state. We know that for any
finite and even $N$ the total spin $s$ of the ground state depends on the value of the coupling
$\phi$. In particular, ground state has spin zero for $\frac{\pi}{2}\leq\phi\leq\pi$ and 
spin $s$ for $\frac{\pi}{2(s+1)}\leq\phi\leq\frac{\pi}{2s}$.
The quasimomenta are known and they are all real. We find that beyond these intervals there is a region
of $\phi$ values where these solutions  become complex. To see this feature
we remind that due to the relation
$M=\frac{N}{2}-s$ and (\ref{ps}) the overall ground state is in the
$M=2$ sector for
\beq \label{gs2}
\frac{\pi}{N-2}\leq\phi\leq\frac{\pi}{N-4}.
\eeq
In particular for $N=6, 8, 10$ these intervals in coupling will be $[\frac{\pi}{4},
\frac{\pi}{2}],[\frac{\pi}{6},\frac{\pi}{4}] \mbox{~and~} [\frac{\pi}{8},\frac{\pi}{6}]$
, respectively. Following this state we see that at certain points outside the above intervals
the quasimomenta become complex. From the choice of Bethe numbers
(\ref{30}) we see that this state will become complex c-solution and the transformation
of this real state into string will happen in the $E=0$ point. Generally, we can find points
in coupling constant where the complex solutions will become real. In the transition points 

\beq \label{uvj1}
\cos\phi=\cos k_{r} \quad\quad\mbox{c-strings}
\eeq

\beq \label{uvj2}
\cos\phi\left(1-\frac{2}{N}\right)=\cos k_{r}\quad\quad\mbox{s-strings}.
\eeq

For c-strings this is satisfied for
\beq \label{fi}
\phi=\frac{\pi}{2(N-1)}[2, ..., N-4, N-2]\quad\quad 0\leq\phi\leq\frac{\pi}{2}
\eeq
\beq \label{fi1}
\phi=\frac{\pi}{2(N-1)}[N, ..., 2N-6, 2N-4]\quad\quad \frac{\pi}{2}\leq\phi\leq\pi.
\eeq
Now we see that the state, which is the overall ground state in interval 
(\ref{gs2}), after a change of $\frac{\pi}{(N-1)(N-2)}$ in $\phi$ becomes c-string. It is interesting 
that the points (\ref{fi}) and (\ref{fi1}) correspond to the points (\ref{24}) where the
the representation theory is no more isomorphic to $SU(2)$ and where several 
multiplets merge together forming indecomposable combinations \cite{A1}. The points where s-strings
disappear
are not of this form because of the correction factor $\left(1-\frac{2}{N}\right)$ in
(\ref{uvj2}).

Crossing of energy levels of complex solutions with the change of coupling constant is one
of the features of this model that make it different from the XXZ model. This is illustrated
on Fig.5 and Fig.6 on which we follow energy levels of $M=2$ strings for both models and for 
number of sites $N=15$. The number of strings is the same in both models. However, in
XXZ model there are complex solutions with $\pm k_{r}$ which are degenerated. Presence
of a term linear in $\phi$ in equations (\ref{bet}), which can be interpreted as a coupling
constant dependent toroidal twist in XXZ model, removes this degeneracy and causes crossing of
energy levels.

\section{Real quasimomenta and completeness of solutions to Bethe equations}
\setcounter{equation}{0}

Now we want to enumerate all real solutions of Bethe equations in two-particle sector.
 We start again from equations  (\ref{tps}) and (\ref{tps1}).
 After manipulating their difference and sum we obtain the following equations
for $k=k_{1}+k_{2}$ and $k_{1}-k_{2}$
\beq \label{pp}
\frac{k}{2}=\frac{1}{N}[\pi(\lambda_{1}+\lambda_{2})-\phi(2-N)]
\eeq
\beq \label{sin}
\frac{\sin (\frac{(k_{1}-k_{2})}{2}(\frac{N}{2}-1))}{\sin (\frac{(k_{1}-k_{2})}{2}\frac{N}{2})}
=\frac{\cos \frac{k}{2}}{\cos\phi}
\qquad\lambda_{1}+\lambda_{2}\quad \mbox{odd}
\eeq
\beq \label{cos}
\frac{\cos (\frac{(k_{1}-k_{2})}{2}(\frac{N}{2}-1))}{\cos (\frac{(k_{1}-k_{2})}{2}\frac{N}{2})}
=\frac{\cos \frac{k}{2}}{\cos\phi}
\qquad\lambda_{1}+\lambda_{2}\quad \mbox{even}.
\eeq

We have to notice that not all $2N-1$ different values of $I=\lambda_{1}+\lambda_{2}$
give different solutions.
Changing $I$ by $N$ is equivalent to changing one quasimomentum by $2\pi$ which results in
change of sign of right hand sides of (\ref{sin}) and (\ref{cos}). 
This reduces the number of possible values of $I$ to $N$, which in turn can be
chosen to give positive values of right hand sides of  (\ref{sin}) and (\ref{cos}).
The left hand sides of the equations (\ref{sin}) and (\ref{cos}) are periodic functions. 
Thus in principle, for each of $N$ different  fixed values of the right hand side one can
count number of solutions by counting number of intersections. These periodic functions are
given on Fig.7 for $N=9$ and $N=10$. For $N$ odd we find that the number of
intersections is 
\beq \label{noe}
N_{o}^{e}=\frac{N-1}{2}\theta\left(\frac{\cos\frac{k}{2}}{\cos\phi}-1\right)+
\frac{N-3}{2}\theta\left(1-\frac{\cos\frac{k}{2}}{\cos\phi}\right)\quad\mbox{I even}
\eeq
\beq \label{noo}
N_{o}^{o}=\frac{N-1}{2}\theta\left(\frac{\cos\frac{k}{2}}{\cos\phi}-1+\frac{2}{N}\right)+
\frac{N-3}{2}\theta\left(1-\frac{2}{N}-\frac{\cos\frac{k}{2}}{\cos\phi}\right)\quad\mbox{I odd}
\eeq
while for $N$ even it is
\beq \label{nee}
N_{e}^{e}=\frac{N}{2}\theta\left(\frac{\cos\frac{k}{2}}{\cos\phi}-1\right)+
\frac{N-2}{2}\theta\left(1-\frac{\cos\frac{k}{2}}{\cos\phi}\right)\quad\mbox{I even}
\eeq
\beq \label{neo}
N_{e}^{o}=\frac{N-2}{2}\theta\left(\frac{\cos\frac{k}{2}}{\cos\phi}-1+\frac{2}{N}\right)+
\frac{N-4}{2}\theta\left(1-\frac{2}{N}-\frac{\cos\frac{k}{2}}{\cos\phi}\right)\quad\mbox{I
odd}.
\eeq
The number of complex solutions is just equal to the second $\theta$ function in the expressions above.
If we take into account that for $N$ even there are $\frac{N}{2}$ even and $\frac{N}{2}$
odd values of $I$ and for $N$ odd $\frac{N-1}{2}$ even values of $I$ and $\frac{N+1}{2}$
odd values of $I$, we find that for any number of sites $N$ and coupling constant $\phi$ 
there are 
\beq \label{broj}
\frac{N(N-1)}{2}
\eeq
solutions to the Bethe equations in the two-particle sector. Among these solutions there are 
$N$ previously identified non-highest weight states, which have one of quasimomenta
equal to $\phi$. Finally, we obtain 
\beq \label{broj1}
{N \choose 2}-{N \choose 1}
\eeq
highest weight Bethe states.

\section{Conclusion}
\setcounter{equation}{0}

In this paper we have investigated complex solutions of the Bethe equations for the 
$M=2$ sector of the $SU(2)_{q}$ invariant closed spin chain for arbitrary number of
sites and coupling constant. We find that some properties of these solutions are  
similar to the properties of complex solutions for the normal XXZ chain and 
some of them are not. In particular we find the number of complex solutions
both for finite $N$ and in thermodynamic limit. This number differs from the number
predicted by the string conjecture for a finite number of solutions. However, if 
$\phi\rightarrow 0$ sufficiently fast compared to $\frac{1}{N}$, the number of exceptions
becomes infinite and goes as $\sqrt{N}$. These properties are essentially the same
as in the case of XXZ chain. One can follow evolution of string solutions and their
disappearance with decreasing 'coupling strength' $|\cos\phi|$. One class of strings
(c-strings) disappears in points where the representation of the $SU(2)_{q}$ is no more
isomorphic to $SU(2)$. By increasing number of sites for fixed coupling constant, 
strings of the other class turn into real quasimomenta that can be found by iterating
Bethe equations with two identical Bethe numbers. This is again violation of the 
string conjecture. The overall ground state, which is always real,
is in the $M=2$ sector for coupling constant 
$\frac{\pi}{N-2}\leq\phi\leq\frac{\pi}{N-4}$. This state becomes complex solution for
$\phi\leq\frac{\pi}{N-1}$. The energy dependence of string solutions shows some simple
features. The strings are found in a narrow energy band and are located on the top of the spectrum near
the antiferromagnetic point and on the bottom near the ferromagnetic point. Their
energy levels cross each other with the change of coupling constant, which is not the case
for the XXZ chain. Finally, we find the number of real solutions. The number of
all solutions of the Bethe equations is ${N \choose 2}$. 
Among these we identify $N$ non-highest weight states. That leads to the ${N \choose 2}
-{N \choose 1}$ highest weight Bethe states.

%


\begin{figure}
\caption{This figure shows dependence of real part of complex s-solutions on numbers of sites
$N$ for $\phi=0.32$. It clearly illustrates transmutation of one complex solution in real solution
(two quasimomenta) for a given critical $N$. These two real quasimomenta correspond to same Bethe 
numbers and are obtained by numerical iteration of equations (\ref{tps}) and (\ref{tps1}).
}

\end{figure}

\begin{figure}
\caption{Spectrum of $M=2$ sector for $N=6$. Energies of both real and complex solutions
of the Bethe equations are plotted.
}

\end{figure}

\begin{figure}
\caption{Same as Fig.~2 but for $N=8$.}

\end{figure}

\begin{figure}
\caption{Same as Fig.~2 but for $N=10$.}

\end{figure}

\begin{figure}
\caption{Dependence of energies of  complex solutions for $SU(2)_{q}$ invariant spin 
chain and number of sites $N=15$ on coupling constant $\phi$.
}

\end{figure}

\begin{figure}
\caption{ Dependence of energies of  complex solutions for XXZ spin 
chain and number of sites $N=15$ on coupling constant $\Delta$, which corresponds
to $\cos\phi$.}

\end{figure}

\begin{figure}
\caption{Left hand sides of equations (\ref{sin}) and (\ref{cos}) as a functions
of $k_{1}-k_{2}$ for $N=9$ and $N=10$. 
}

\end{figure}

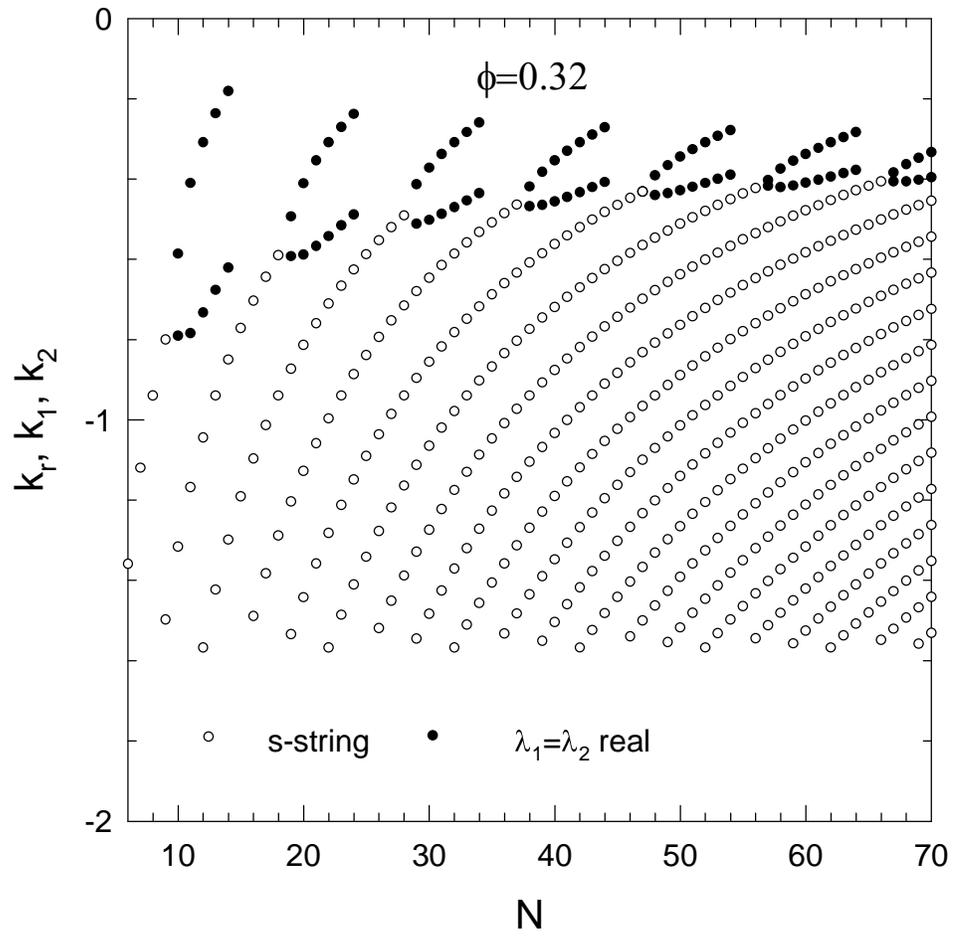
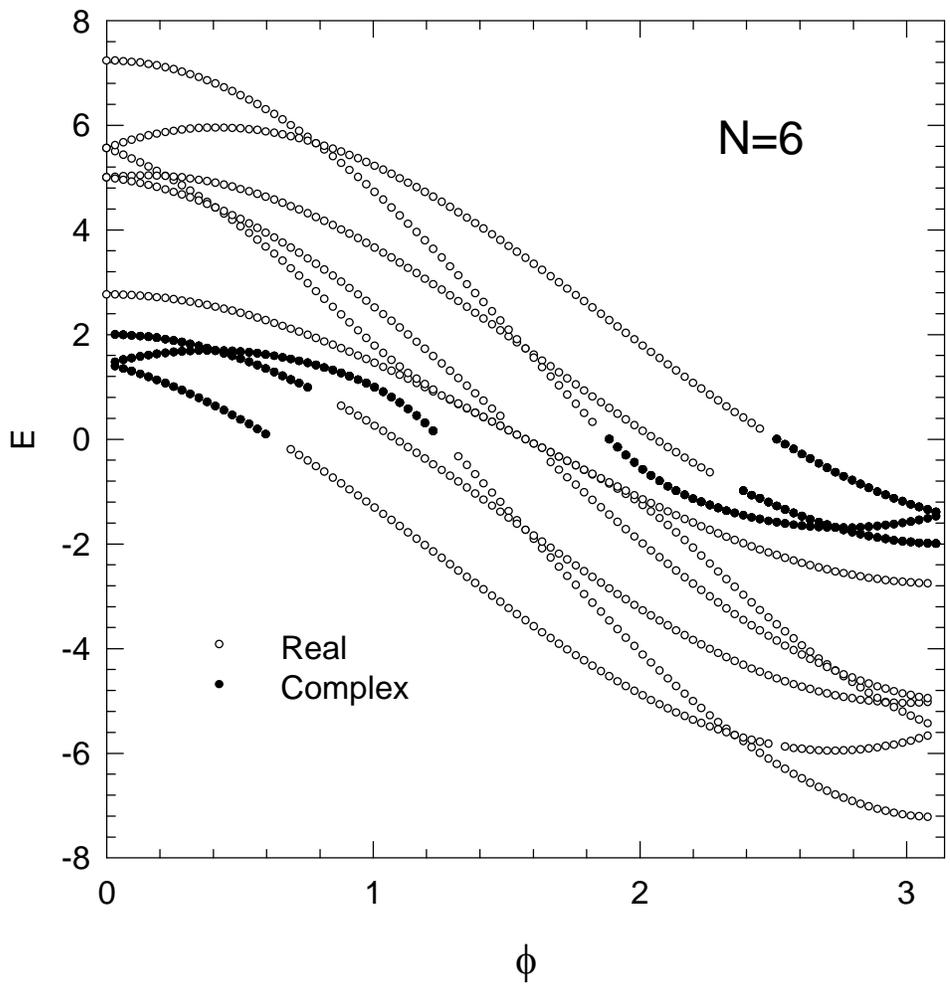
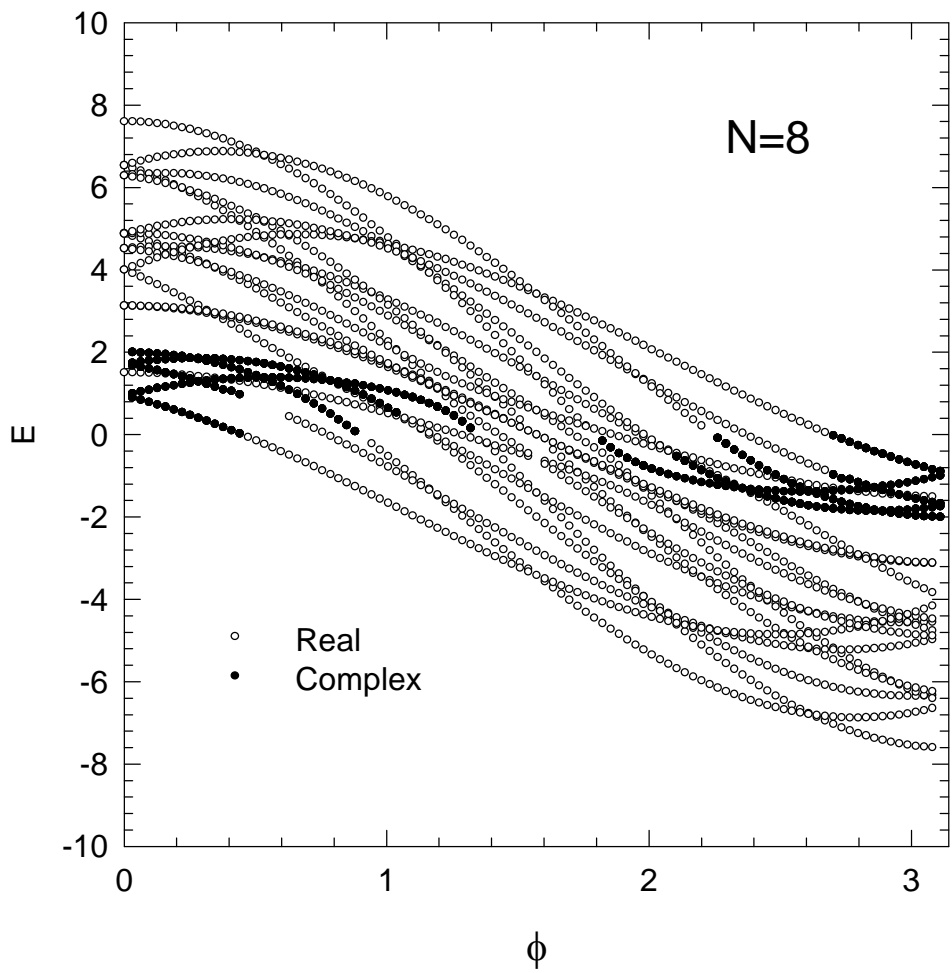
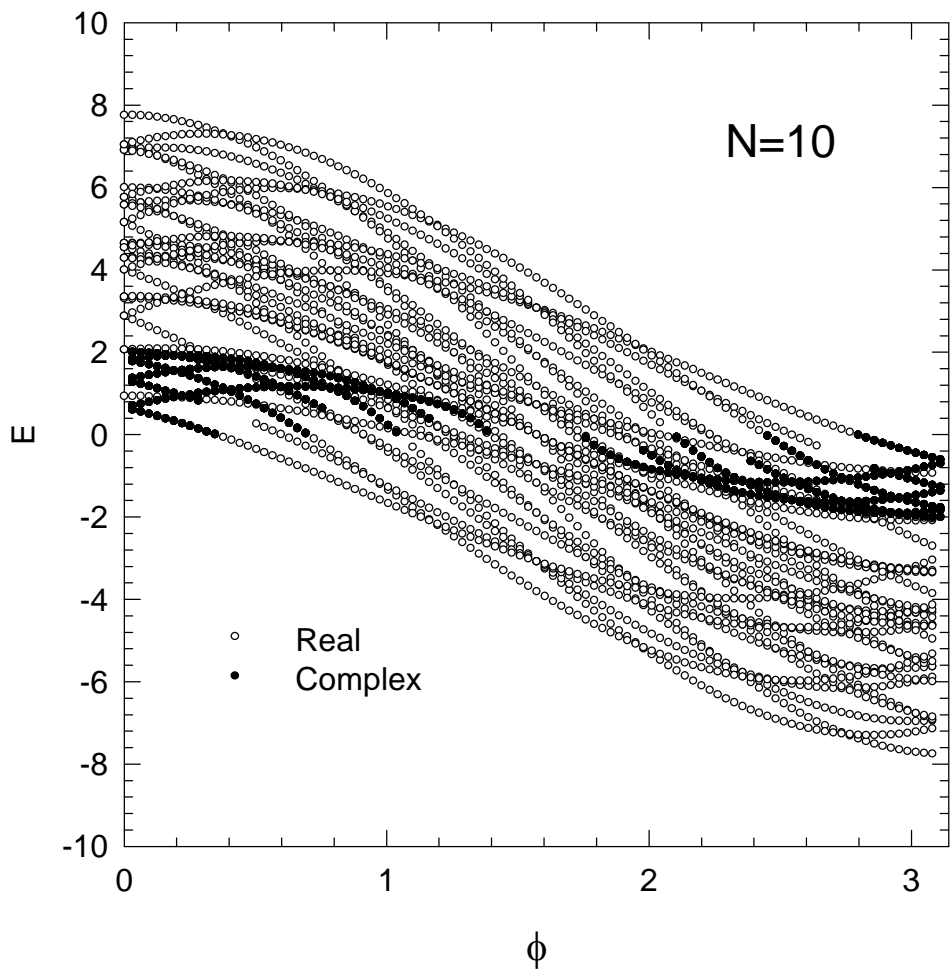
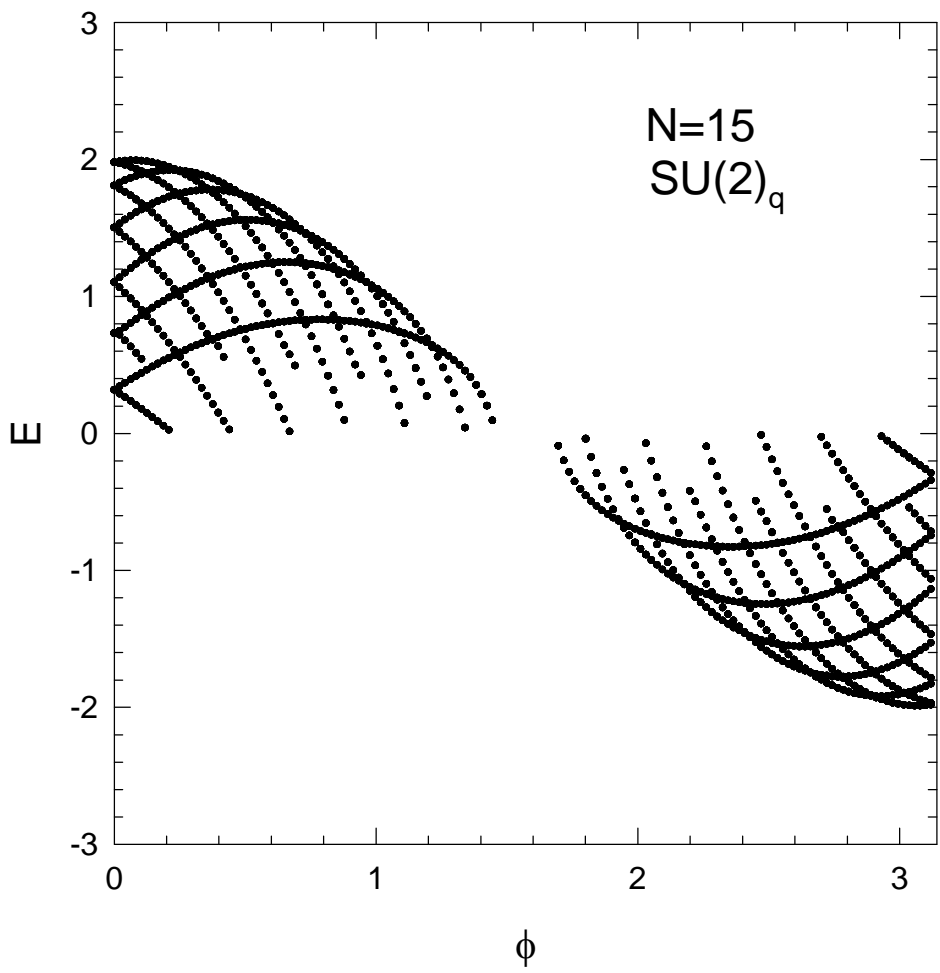
\end{document}